\documentclass[conference,compsoc]{IEEEtran}
\IEEEoverridecommandlockouts
\usepackage{amsmath,amssymb,amsfonts}
\usepackage{url}
\usepackage{tabularx}
\newcolumntype{Y}{>{\centering\arraybackslash}X}
\usepackage{enumitem}
\usepackage[skip=0pt]{caption}
\usepackage{setspace}
\usepackage{multirow}
\usepackage{amssymb}
\usepackage{booktabs}
\usepackage{bbm}
\RequirePackage{subcaption}

\usepackage{algorithmicx}
\usepackage[ruled,vlined,linesnumbered]{algorithm2e}
\algnewcommand\algorithmicinput{\textbf{Input:}}
\algnewcommand\algorithmicoutput{\textbf{Output:}}

\usepackage{tcolorbox}
\usepackage{listings}
\usepackage{xcolor}

\definecolor{darkgreen}{rgb}{0.1, 0.2, 0.1}

\usepackage{graphicx}
\usepackage{textcomp}
\usepackage{xcolor}
\usepackage{hyperref}

\usepackage{fvextra}
\usepackage{amssymb} 

\DefineVerbatimEnvironment{CustomVerbatim}{Verbatim}{
    breaklines=true,               
    breaksymbolleft=$\hookrightarrow$,             
    fontsize=\small,               
    baselinestretch=0.85            
}

\usepackage{color}

\def\BibTeX{{\rm B\kern-.05em{\sc i\kern-.025em b}\kern-.08em
    T\kern-.1667em\lower.7ex\hbox{E}\kern-.125emX}}
\begin{document}

\title{\textsc{Insight-LLM}: LLM-enhanced Multi-view Fusion in Insider Threat Detection}


\author{\IEEEauthorblockN{Chengyu Song}
\IEEEauthorblockA{\textit{Systems Engineering Institute} \\
\textit{Academy of Military Sciences}\\
Beijing, China \\
songchengyu@alumni.nudt.edu.cn}
\and
\IEEEauthorblockN{Jianming Zheng$^{\star}$}
\IEEEauthorblockA{\textit{Laboratory for Advanced Computing} \\ {and Intelligent Engineering}\\
Wuxi, China \\
zhengjianming12@nudt.edu.cn}
}


\maketitle

\begin{abstract}
Insider threat detection (ITD) requires analyzing sparse, heterogeneous user behavior.
Existing ITD methods predominantly rely on single-view modeling, resulting in limited coverage and missed anomalies. 
While multi‑view learning has shown promise in other domains, its direct application to ITD introduces significant challenges: scalability bottlenecks from independently trained sub‑models, semantic misalignment across disparate feature spaces, and view imbalance that causes high‑signal modalities to overshadow weaker ones.
In this work, we present Insight-LLM, the first modular multi‑view fusion framework specifically tailored for insider threat detection. 
Insight-LLM employs frozen, pre‑trained encoders to extract high‑quality embeddings for each behavioral view, avoiding overfitting and reducing training overhead. 
Then, Dedicated Qformers align each embedding into the LLM’s semantic manifold for deep cross‑modal fusion.
Subsequently, a cross‑view attention layer then re‑weights embeddings to amplify weak threat cues and suppress noise.
Finally, fused embeddings are tuned via multi-view fusion for robust ITD under severe class imbalance.
Experiments on two real‑world datasets show Insight-LLM outperforms baselines, achieving state‑of‑the‑art detection with low latency and parameter overhead.
\end{abstract}


\section{Introduction}
\label{Introduction}

Insider threats pose a significant risk in today’s digital landscape. 
Trusted users, whether acting maliciously or inadvertently, can exploit their access to steal, alter, or damage critical assets.
According to the Ponemon Institute’s 2023 Cost of Insider Threats Global Report, organizations require over 80 days on average to contain an insider incident, and the 2025 edition estimates annual losses of \$17.4 million per organization~\cite{cost_of_ITD_2023, cost_of_ITD_2025}. 
These high stakes have driven extensive research into Insider Threat Detection (ITD), with state-of-the-art approaches modeling user behavior to flag anomalies~\cite{DBLP:journals/compsec/FeiZZGFLWC25,DBLP:conf/ipccc/SunY22,DBLP:conf/ccs/Du0ZS17}.

Despite these advances, most ITD methods focus on a single behavioral dimension, such as temporal action sequences~\cite{DBLP:conf/aaai/TuorKHNR17a}, textual content~\cite{DBLP:conf/icbar/WangZXF24}, or the topology of user-entity graphs~\cite{DBLP:journals/tifs/CaiWXLZLY24}. 
In practice, insiders camouflage malicious activities amid vast volumes of legitimate behavior, yielding sparse and heterogeneous anomaly signals~\cite{DBLP:journals/compsec/FeiZZGFLWC25}.

\begin{figure}[t]
	\centerline{\includegraphics[width=\columnwidth]{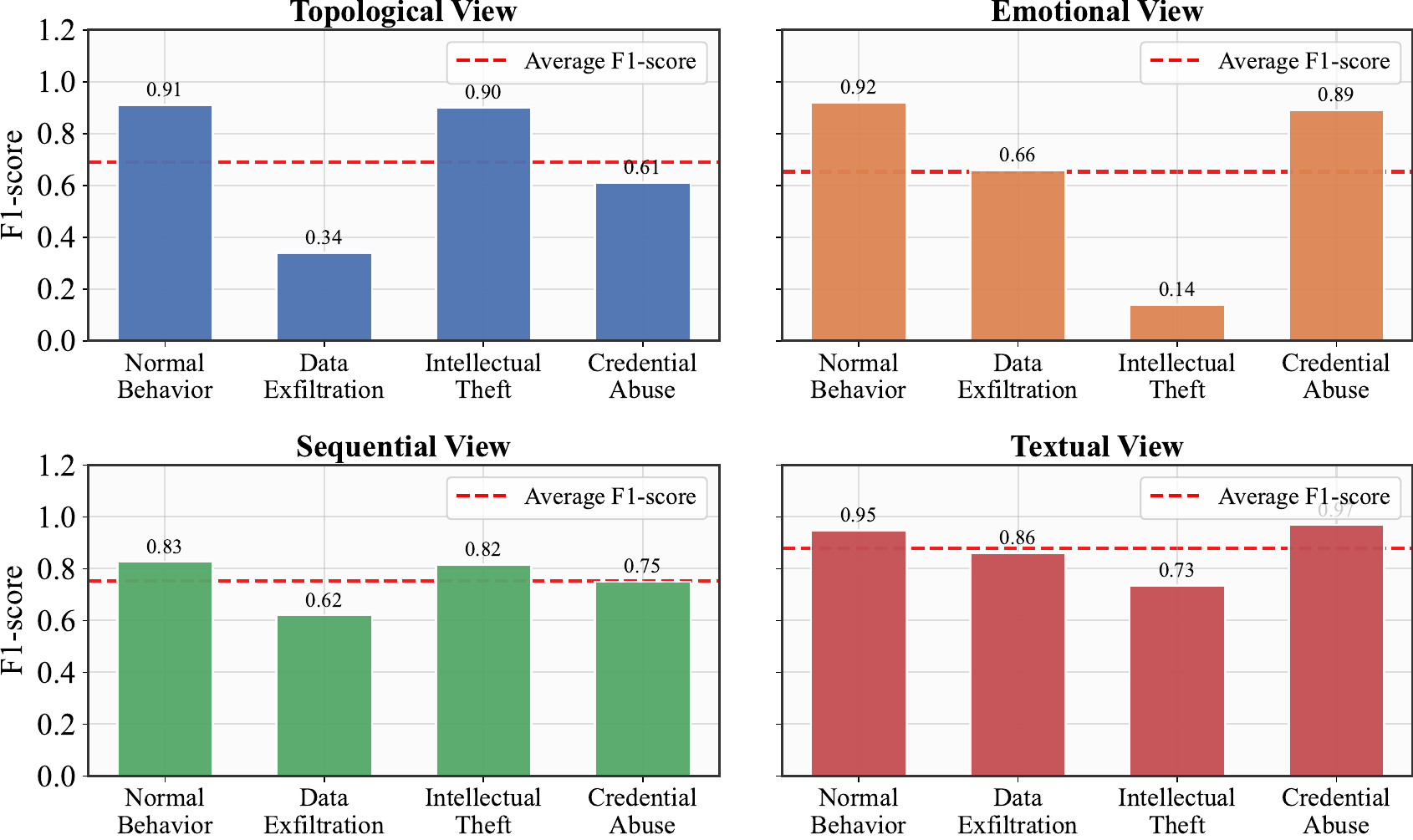}}
	\caption{Performance of single-view analysis across different threat types.}
	\label{pilot_experiment}
\end{figure}

\vspace{0.5em}
\noindent \textbf{\textit{Pilot Experiments}}
To examine the limitations of single-view ITD methods, we conduct experiments across the four most commonly used behavioral views in this domain, i.e., text, topology, sequence, and sentiment.
To isolate the effect of each view, we extract embeddings with dedicated unsupervised encoders.
In detail, RoBERTa~\cite{DBLP:journals/corr/abs-1907-11692} for text and sentiment, TS2Vec~\cite{DBLP:conf/aaai/YueWDYHTX22} for sequence data, and Node2Vec~\cite{DBLP:conf/kdd/GroverL16} for topology, without incorporating any cross-view information. 
A linear classifier is then applied for threat prediction. 
As shown in Fig.~\ref{pilot_experiment}, single-view models either fail to identify certain threat categories or perform poorly overall. 
These results reveal two fundamental characteristics of ITD: 
(1) threat signals are inherently sparse and broadly distributed, limiting the effectiveness of any single view; and 
(2) different views capture distinct aspects of threat behavior, exhibiting complementary detection capabilities. 
Together, these insights underscore the need for a unified framework that effectively integrates the strengths of multiple behavioral views.

\vspace{0.5em}
\noindent \textbf{\textit{Challenges}}
Despite the success of multi-view learning in domains such as deep clustering and representation learning~\cite{DBLP:journals/pami/LiJHWZLMZ25,DBLP:conf/aaai/0001C0WN25}, applying these techniques to ITD presents unique challenges due to the nature of security-critical, behavior-sparse data. Specifically, three key challenges arise:

\begin{itemize}[leftmargin=*,nosep]
    \item \textbf{Ever-Increasing Parameters Vs. Resource-Intensive Deployments}
    Traditional multi-view fusion methods often train a separate sub-model for each view and then apply a late-fusion strategy (e.g., isolation forest~\cite{DBLP:journals/compsec/FeiZZGFLWC25}). This design is manageable when the number of views is small and relatively homogeneous. However, in ITD scenarios, user behavior data spans a broader range of heterogeneous modalities—including distance-aware topological structures, temporal activity sequences, and textual communications. The number and diversity of views lead to a rapid increase in parameter size and inference cost, rendering such late-fusion approaches resource-intensive and impractical for real-time or large-scale deployment.
    
    \item \textbf{Aligned Modalities Vs. Disjoint Behaviors}
    Fusion strategies that assume temporal and semantic alignment between views are effective in tasks like audio-visual event detection, where modalities co-occur and share consistent semantics. In insider threat detection, views such as email content, access logs, and network topology are generated independently, with differing time granularities and weak cross-view correlations. This misalignment makes it difficult to associate subtle threat cues scattered across views, limiting the effectiveness of conventional multi-view fusion approaches.
    
    \item \textbf{Matthew Effect VS. Vital Few Matters}
    In typical multi-view tasks, class distributions are often balanced, enabling models to learn all categories relatively evenly. In contrast, ITD faces extreme class imbalance: benign behaviors dominate, while actual threats are both rare and diverse. This skew leads to a Matthew Effect during training, where models overfit to frequent benign patterns and overlook the sparse but critical threat instances. Consequently, conventional multi-view methods struggle to identify the “vital few” anomalies that truly matter for security.
\end{itemize}

\vspace{0.5em}
\noindent \textbf{\textit{Approach and Contributions}}.
In this paper, we propose Insight-LLM, a unified yet modular framework specifically designed to address the core challenges of applying multi-view learning to ITD.
To address \textbf{Challenge 1}, Insight-LLM adopts a per-view modular design. 
Each behavioral view, such as text content, action sequences, or graph-based user relations, is encoded by a frozen, pre-trained encoder, ensuring high-quality representations without increasing training overhead. 
This avoids model duplication and reduces both parameter count and inference latency, enabling efficient scaling as the number of views grows.
To mitigate \textbf{Challenge 2}, each view’s representation is passed through a dedicated Qformer~\cite{DBLP:conf/icml/0008LSH23}, a view-specific cross-attention structure that captures internal dependencies and projects the feature into the semantic space of a pre-trained LLM.
Inspired by CLIP-style modality alignment~\cite{DBLP:conf/icml/RadfordKHRGASAM21,DBLP:conf/nips/NarasimhanRD21}, this design maps heterogeneous behavioral signals into a shared semantic manifold, shaped by large-scale language pertaining.
In this space, semantically related behaviors from different views naturally cluster, enabling consistent comparison across heterogeneous structures.
By leveraging semantic proximity, Insight-LLM amplifies weak cross-modal threat cues and enhances anomaly separation, without the need to retrain large fusion networks.
To tackle \textbf{Challenge 3}, we introduce a cross-view attention module that dynamically re-weights each view based on its contextual relevance. 
This mechanism amplifies weak but semantically consistent threat cues, especially from low-signal or underrepresented views, while attenuating the influence of noisy or dominant modalities.
Finally, the fused representation is concatenated with a task-specific prompt and passed to the LLM, where we perform multi-view fusion tuning via LoRA-based lightweight fine-tuning. 
Together, these components enable Insight-LLM to integrate multi-view behavioral signals in a scalable, semantically aligned, and dynamically balanced manner, meeting the unique demands of insider threat detection.
The main contributions of this paper are summarized as follows:
\begin{itemize}[leftmargin=*,nosep]
    \item To the best of our knowledge, Insight-LLM is the first multi-view fusion framework tailored for insider threat detection, which unifies heterogeneous behavioral signals within a shared semantic space to amplify weak supervisory cues indicative of malicious insider activities.

    \item We design a modular per-view Qformer architecture that aligns multi-view data into a unified semantic embedding space, enabling deep fusion without incurring model duplication or scalability bottlenecks.

    \item To address view imbalance and noise heterogeneity, we introduce a cross-view attention mechanism that dynamically re-weights view contributions based on contextual anomaly signals, allowing weak but consistent threat cues to be effectively amplified.

    \item Extensive experiments on two insider threat datasets show that Insight-LLM effectively integrates multi-view data to capture weak threat signals, achieving state-of-the-art detection performance.
\end{itemize}
\section{Related Work}
\label{relatedwork}

\subsection{Log-based Insider Threat Detection}

Insider threat detection has gained significant research attention over the last decade due to its critical role in cybersecurity.
Existing approaches fall into two categories: rule-based matching and machine-learning-based methods.
Rule-based matching is among the earliest and most widely adopted approaches.
For example, early tools such as Addamark LMS~\cite{DBLP:conf/lisa/Sah02} and Splunk~\cite{hanley_2011} leveraged rule-matching mechanisms to flag suspicious behavior.
This rule-matching capability allowed analysts to audit logs and trigger precise alerts for predefined threat behaviors.
However, because these systems rely on fixed rules, they struggle to adapt to emerging or previously unseen threats.
To mitigate this limitation, Milajerdi et al.~\cite{DBLP:conf/sp/MilajerdiGESV19} incorporated MITRE ATT\&CK TTPs~\cite{mitreMITREATTampCKxAE} to map low-level audit logs to higher-level attack stages, thus broadening detection scope.
Although effective for specific, well-understood threats, such methods require extensive domain expertise and offer limited adaptability to dynamic threat landscapes~\cite{DBLP:journals/tnsm/XiaoYZWCL23}.

To address the generalization shortcomings of static rules, research efforts have increasingly turned to machine-learning-based approaches.
Motivated by the temporal dependencies in user behavior, one mainstream strategy models audit logs as sequential data.
For example,
DeepLog~\cite{DBLP:conf/ccs/Du0ZS17} uses LSTMs to flag deviations in log sequences. 
Tuor et al.~\cite{DBLP:conf/aaai/TuorKHNR17a} proposed a stacked LSTM model that quantifies anomaly scores using the log‑likelihood of observed user activity sequences. 
Complementing these, Yuan et al.~\cite{DBLP:conf/bigdataconf/YuanZWL19} modeled event types and timestamps via hierarchical temporal point processes to score behavioral anomalies.
More recently, graph-based representations have been adopted to capture complex user–entity relationships and structural anomalies. 
For instance, 
Oka et al.~\cite{oka2004eigen} built user–system bipartite graphs to detect latent threats from anomalous interactions. 
Log2vec~\cite{DBLP:conf/ccs/LiuWZJXM19} embeds event semantics in a heterogeneous graph and applies unsupervised clustering to surface abnormal behavior patterns.

However, most existing machine-learning-based ITD methods either rely on a single isolated view or prioritize a dominant modality while loosely incorporating others.
This results in either overlooking dispersed threat cues or relying on naïve fusion strategies that overshadow complementary signals.
In contrast, our Insight-LLM projects multi-view data into a shared semantic space via cross-view attention, enabling adaptive fusion and uncovering subtle, otherwise-missed cross-modal anomalies.

\subsection{Multi-View Representation Fusion}
Multi-view representation learning seeks to capture complementary information from heterogeneous sources, with fusion techniques aiming to integrate these into a compact, unified representation~\cite{DBLP:journals/tkde/LiYZ19}. 
This paradigm has been widely explored within neural network architectures, including autoencoders, convolutional neural networks (CNNs), and recurrent neural networks (RNNs).
For deep autoencoders, 
Ngiam et al.~\cite{DBLP:conf/icml/NgiamKKNLN11} proposed a bimodal architecture that learns shared representations by concatenating audio and video signals, projecting them into a common latent space. 
Cadena et al.~\cite{DBLP:conf/rss/CadenaDR16} applied this concept to RGB and depth data fusion for missing data recovery.
In the context of CNNs, 
Su et al.~\cite{DBLP:conf/iccv/SuMKL15} introduced a multi-view CNN that processes 2D projections of 3D objects through shared convolutional layers, followed by view pooling to aggregate features. 
Feichtenhofer et al.~\cite{DBLP:conf/cvpr/FeichtenhoferPZ16} further explored spatiotemporal fusion strategies to enhance video-based action recognition.
With respect to RNNs, 
Cho et al.~\cite{DBLP:conf/emnlp/ChoMGBBSB14} developed an encoder–decoder model that learns conditional mappings between variable-length multimodal sequences. 
Karpathy and Li.~\cite{DBLP:journals/pami/KarpathyF17} proposed a multimodal RNN for generating region-level image descriptions, while 
Chen and Zitnick~\cite{DBLP:journals/corr/ChenZ14a} investigated bidirectional mappings between images and textual descriptions. 
Venugopalan et al.~\cite{DBLP:conf/iccv/VenugopalanRDMD15} extended this to an end-to-end sequence model for video captioning.

While effective in general settings, these approaches often lack robustness in security domains, where signals may be noisy, sparse, or weak. To address this, Insight-LLM introduces adaptive fusion with semantic alignment tailored for threat detection.
\begin{figure*}[htbp]
	\centerline{\includegraphics[width=\textwidth]{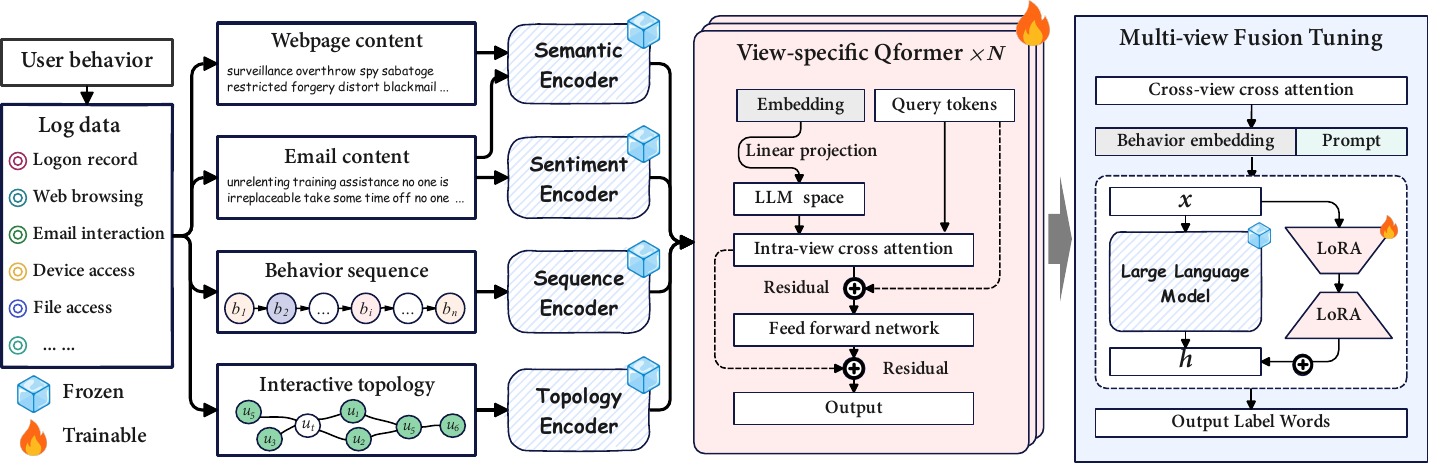}}
	\caption{The workflow of Insight-LLM comprises three stages: (1) partitioning user behavior logs into distinct views and encoding each with a frozen pretrained model to produce feature embeddings; (2) projecting these embeddings into the LLM’s semantic manifold via view‑specific Qformer modules; and (3) applying cross‑view attention to dynamically re‑weight the aligned embeddings and then tuning the LLM for multi‑view fusion.}
	\label{frame}
	\label{overall_framework}
\end{figure*}
\section{Approach}

\subsection{Task Definition}
Consider a sequence of behavioral logs $\mathcal{L}=\{l_1, l_2, \cdots, l_n\}$ generated by an internal user within an information system.
Each log entry $l_i$ contains event-level information such as timestamp, event type, and various behavioral attributes.
Events span multiple types, each contributing different semantic cues.
The goal of ITD is to detect malicious behaviors within $\mathcal{L}$ by learning a model $\mathcal{M}$ that identifies a subset of events $\mathcal{L}_m \subseteq \mathcal{L}$ as suspicious.
Single-view ITD approaches operate on a specific feature view and make predictions, formulated as:
\begin{equation}
    \mathcal{L}_m = \{l_i \in \mathcal{L} \mid \mathcal{M}_v(l_i) = \text{malicious}\},
\end{equation}
where $\mathcal{M}_v$ is a model trained on features extracted from a single behavioral view.
In contrast, multi-view framework models $\mathcal{V}$ semantic views, where each view corresponds to a distinct latent perspective.
Instead of assigning views based on event types, we construct each view by projecting all logs into view-specific latent spaces via dedicated mapping functions:
\begin{equation}
    \mathcal{X}^{(v)} = \{\phi_v(l_i) \mid l_i \in \mathcal{L}\}, \quad v = 1,\cdots,\mathcal{V},
\end{equation}
where $\phi_v(\cdot)$ extracts features relevant to the $v$-th semantic view from each event, regardless of its type.
%
%
The detection model then aligns and integrates representations from all views to perform joint prediction:
\begin{equation}
    \mathcal{M}(\mathcal{X}^{(1)}, \cdots, \mathcal{X}^{(\mathcal{V})}) = F\left( \bigoplus_{v=1}^{\mathcal{V}} Q_v(\mathcal{X}^{(v)}) \right),
\end{equation}
where $Q_v$ transforms each view into a unified semantic space, $\bigoplus$ denotes a cross-view fusion operator (e.g., attention, gating), and $F$ is the final classification function.
Based on per-log predictions from the joint multi-view model, the malicious subset $\mathcal{L}_m$ is determined as:
\begin{equation}
    \mathcal{L}_m=\{l_i\in\mathcal{L}|F\left( \bigoplus_{v=1}^{\mathcal{V}} Q_v(\phi_v(l_i))\right)=\text{malicious}\}
\end{equation}

\subsection{Multi-view Feature Extraction}
To detect insider threats concealed within overwhelming volumes of benign activity, we decompose user behavior into four complementary views\textemdash textual, sentiment, sequential, and topological\textemdash each of which is encoded by a specialized, dedicated model.
The textual view captures message semantics (e.g., emails, web content) to reveal intent or policy violations, and is encoded with a vanilla RoBERTa model~\cite{DBLP:journals/corr/abs-1907-11692}.
The sentiment view captures affective cues, such as frustration or hostility, by encoding text with a sentiment‑adapted RoBERTa model (twitter‑roberta‑base‑sentiment‑latest)~\cite{camacho-collados-etal-2022-tweetnlp}.
The sequential view models temporal dynamics in user actions to detect anomalous routines, applying TS2Vec~\cite{DBLP:conf/aaai/YueWDYHTX22} for unsupervised time‑series encoding.
Following~\cite{DBLP:conf/iscc/HuangZLLWY21}, we bucket each timestamp’s hour into an integer in $\{0,\cdots,23\}$ and comnpute:
\begin{equation}
    c_i=\text{type}(l_i)\times24+\text{time}(l_i),
\end{equation}
where $\text{type}(l_i)$ is the activity type ID and time $\text{time}(l_i)$ its hourly slot.
The topological view encodes user–resource and user–user interaction graphs using Node2Vec~\cite{DBLP:conf/kdd/GroverL16}, whose biased random walks yield structure-preserving embeddings to reveal lateral movement or collusion.

These encoders share three key characteristics that make them particularly suitable for our scenario. 
First, they are \textbf{lightweight}, which avoids excessive computational overhead during feature extraction and facilitates scalable deployment. 
Second, they are \textbf{frozen}, meaning their parameters remain unchanged during training; this ensures stable and consistent representations even under severe class imbalance. 
Third, they are \textbf{unsupervised}, thus eliminating the need for labeled data or task-specific tuning, which is particularly advantageous in domains where annotation is costly or infeasible. 
Together, these properties enable robust and view-consistent feature extraction, providing a solid foundation for downstream alignment and multimodal fusion.
We denote the extracted multi-view behavioral features as:
\begin{equation}
    \begin{aligned}
    &\mathcal{X}_{\text{input}} = [\mathcal{X}_{\text{text}}, \mathcal{X}_{\text{sent}}, \mathcal{X}_{\text{seq}}, \mathcal{X}_{\text{topo}}], \text{where}\\
    & \quad
        \begin{aligned}[t]
            \mathcal{X}_{\text{text}} &= \text{RoBERTa}_v(M_{\text{text}}), && \in \mathbb{R}^{L_t\times d_t} \\
            \mathcal{X}_{\text{sent}} &= \text{RoBERTa}_s(M_{\text{sent}}), && \in \mathbb{R}^{L_s\times d_s} \\
            \mathcal{X}_{\text{seq}}  &= \text{TS2Vec}(S),                  && \in \mathbb{R}^{L_q\times d_q} \\
            \mathcal{X}_{\text{topo}} &= \text{Node2Vec}(G),               && \in \mathbb{R}^{L_g\times d_g}
        \end{aligned}
    \end{aligned}
\end{equation}
Where $M_{text}$ and $M_{sent}$ denote the textual and sentiment inputs, respectively; $S$ the event sequence, and $G$ the interaction graph.

\subsection{Cross-view Manifold Alignment}
Despite the rich and complementary features captured by the textual, sentiment, sequential, and topological views, these modalities reside in inherently heterogeneous embedding spaces.
This heterogeneity introduces substantial representation gaps across views, posing a fundamental challenge to unified modeling.
Drawing inspiration from BLIP-2~\cite{DBLP:conf/icml/0008LSH23}, we introduce a dedicated ViewAdapter for each behavioral view, which projects modality-specific features onto a shared latent manifold via query-guided attention.

In detail, let each view yield a sequence of embeddings $\mathcal{X}^{(v)}\in\mathbb{R}^{N_v\times d_v}$, where $N_v$ is the number of tokens and $d_v$ their native dimension.
We first apply a linear projection, followed by layer normalization, to map all embeddings into a common latent manifold of dimension $d$:
\begin{equation}
    \mathcal{X}'^{(v)}=\text{LayerNorm}(\mathcal{X}^{(v)}W_{in}^{(v)})\in\mathbb{R}^{N_v\times d}.
\end{equation}
This step ensures that all views share a common numerical scale and dimensionality, facilitating subsequent alignment and enabling the model to detect or reason over latent cross-view inconsistencies that may signal anomalous insider behavior.
To extract semantically meaningful summaries from each view, we introduce $M$ learnable query tokens $Q_{intra}^{(v)}\in\mathbb{R}^{M\times d}$, which act as view-aware probes to retrieve threat-indicative patterns, and apply an intra-view cross-attention over the normalized embeddings:
\begin{equation}
    Z^{(v)}=\text{Intra-CrossAttn}(Q^{(v)},\mathcal{X}'^{(v)},\mathcal{X}'^{(v)})\in\mathbb{R}^{M\times d},
\end{equation}
where $\text{Intra-CrossAttn}(Q,K,V)$ denotes intra-view multi-head attention. 
Existing single-view ITD models tend to overly rely on high signal-to-noise views, such as textual content, while underutilizing less informative views, like topological or sequential behaviors. 
Our query-guided attention enables each view to identify its most threat-relevant patterns autonomously, amplifying complementary signals and reducing the interference from noisy or redundant features.
To ensure semantic stability and prevent representation drift, we further incorporate a residual connection between the query tokens and the attention output:
\begin{equation}
    Z'^{(v)}=Z^{(v)}+Q^{(v)}.
\end{equation}
The resulting intermediate representation $Z'^{(v)}$ is further refined through a lightweight feed-forward network with another residual connection:
\begin{equation}
    \tilde{Z}^{(v)} = Z'^{(v)} + \text{FFN}^{(v)}(Z'^{(v)}) \in \mathbb{R}^{M \times d}.
\end{equation}
This residual formulation allows the adapter to refine features while preserving the original attention output, facilitating stable and effective learning.
The resulting $\tilde{Z}^{(v)}$ thus provides $M$ aligned tokens per modality, each embedded in the same latent manifold, laying a robust foundation for downstream cross-view fusion.
Each $\tilde{Z}^{(v)} \in \mathbb{R}^{M\times d}$ represents the unified query tokens distilled from modality $v$ via its ViewAdapter.
\begin{equation}
    \begin{aligned}
    &\tilde{Z}_{\text{output}} = [\tilde{Z}_{\text{text}}, \tilde{Z}_{\text{sent}}, \tilde{Z}_{\text{seq}}, \tilde{Z}_{\text{topo}}], \text{where} \quad\\ 
        & \quad
        \begin{aligned}[t]
            \tilde{Z}_{\text{text}} &= \text{ViewAdapter}_{\text{text}}(\mathcal{X}_{\text{text}}) \\
            \tilde{Z}_{\text{sent}} &= \text{ViewAdapter}_{\text{sent}}(\mathcal{X}_{\text{sent}}) \\
            \tilde{Z}_{\text{seq}}  &= \text{ViewAdapter}_{\text{seq}}(\mathcal{X}_{\text{seq}}) \\
            \tilde{Z}_{\text{topo}} &= \text{ViewAdapter}_{\text{topo}}(\mathcal{X}_{\text{topo}})
        \end{aligned} \\
        &\text{and} \quad \tilde{Z}_{\text{text}}, \tilde{Z}_{\text{sent}}, \tilde{Z}_{\text{seq}}, \tilde{Z}_{\text{topo}} \in \mathbb{R}^{M \times d}
    \end{aligned}
\end{equation}

This structure-preserving alignment mechanism is particularly well-suited for multi-view insider threat detection scenarios, where critical threat signals are often subtle and dispersed across diverse and heterogeneous behavioral modalities. 
By effectively bridging representation gaps and harmonizing feature spaces, this approach enables more reliable integration of complementary information, ultimately enhancing the sensitivity and robustness of threat identification.

\begin{algorithm}[t]
    \small
    \caption{Multi-view Insider Threat Detection}
    \label{algorithm}
    \KwIn{
        $\mathcal{L}$: User behavioral logs; 
        $\mathcal{P}$: Prompt text; 
        $\mathcal{V}$: Verbalizer tokens;\\
        $\mathcal{E}$: View-specific encoders; 
        $\mathcal{M}_{LLM}$: LLM with LoRA tuning
    }
    \KwOut{
        $\hat{y}$: Threat label $\in \{0,1\}$
    }

    \BlankLine
    \tcp{1. View-specific Feature Extraction}
    \ForEach{view $v \in \{\text{text}, \text{sent}, \text{seq}, \text{topo}\}$}{
        $\mathcal{X}^{(v)} \gets \mathcal{E}^{(v)}(\mathcal{L})$
    }

    \BlankLine
    \tcp{2. Intra-view Alignment}
    \ForEach{view $v$}{
        $\mathcal{X}'^{(v)} \gets \text{LayerNorm}(\mathcal{X}^{(v)} W_{\text{in}}^{(v)})$\;
        $Z^{(v)} \gets \text{MultiHeadAttn}(Q^{(v)}, \mathcal{X}'^{(v)}, \mathcal{X}'^{(v)})$\;
        $Z'^{(v)} \gets Z^{(v)} + Q^{(v)}$\;
        $\tilde{Z}^{(v)} \gets Z'^{(v)} + \text{FFN}^{(v)}(Z'^{(v)})$
    }

    \BlankLine
    \tcp{3. Inter-view Fusion}
    $\mathbf{Z}_{\text{cat}} \gets [\tilde{Z}_{\text{text}} \Vert \tilde{Z}_{\text{sent}} \Vert \tilde{Z}_{\text{seq}} \Vert \tilde{Z}_{\text{topo}}]$\;
    $\mathcal{F} \gets \text{MultiHeadAttn}(Q_{\text{inter}}, \mathbf{Z}_{\text{cat}}, \mathbf{Z}_{\text{cat}})$

    \BlankLine
    \tcp{4. Prompt-based LLM Prediction}
    $E_{\mathcal{P}} \gets \text{Tokenize}(\mathcal{P})$\;
    $X_{\text{LLM}} \gets [\mathcal{F} \Vert E_{\mathcal{P}}]$\;
    $\mathcal{O} \gets \mathcal{M}_{LLM}(X_{\text{LLM}})$\;
    $h_{\text{final}} \gets \mathcal{O}[-1]$\;
    $\hat{y} \gets \arg \max_{c \in \{0,1\}} \cos(h_{\text{final}}, \mathcal{V}_c)$

    \BlankLine
    \Return $\hat{y}$
\end{algorithm}

\subsection{Multi-view Fusion Tuning}
While each ViewAdapter independently maps its input into the shared semantic manifold, the resulting representations remain isolated across views.
To capture cross-view dependencies essential for identifying subtle threat patterns, we introduce Multi-view Fusion Tuning.
Unlike the intra-view cross-attention within each ViewAdapter, this stage performs \textbf{inter-view cross-attention} using learnable fusion queries to integrate and reweight information across views.
This allows the model to emphasize the most informative view-specific features conditioned on the prediction objective.

Formally, we first concatenate the unified query representations from all views:
\begin{equation}
    \mathbf{Z}_{\text{cat}} = \left[ \tilde{Z}_{\text{text}} \Vert \tilde{Z}_{\text{sent}} \Vert \tilde{Z}_{\text{seq}} \Vert \tilde{Z}_{\text{topo}} \right]
\end{equation}
where $\Vert$ denotes concatenation along the token dimension.
Let $Q_{\text{inter}} \in \mathbb{R}^{N_q \times d}$ denote a set of learnable fusion queries.
The inter-view cross-attention output is computed as:
\begin{equation}
    \mathcal{F} = \text{Inter-CrossAttn}(Q_{\text{inter}}, \mathbf{Z}_{\text{cat}}, \mathbf{Z}_{\text{cat}}) \in \mathbb{R}^{N_q \times d}.
\end{equation}

Theoretically, this cross-view fusion process can be interpreted through the lens of the information bottleneck principle~\cite{DBLP:journals/entropy/LewandowskyB24}, where the fused representation $\mathcal{F}$ is encouraged to retain minimal sufficient information from the multi-view input $\mathbf{Z}_{\text{cat}}$ for predicting behavioral labels $Y$.
This corresponds to minimizing:
\begin{equation}
I(\mathbf{Z}_{\text{cat}}; \mathcal{F}) - \beta I(\mathcal{F}; Y),
\end{equation}
which effectively reduces the conditional entropy $\mathbb{H}(Y|\mathcal{F})$.
This formulation encourages $\mathcal{F}$ to act as a compact, discriminative abstraction that filters out redundant view-specific noise while preserving the threat-relevant signals distributed across views.

To enable classification via the LLM, we concatenate the fused representation $\mathcal{F}$ with a tokenized natural language prompt $\mathcal{P} = \{p_1, ..., p_{L_{\mathcal{P}}}\}$ to form:
\begin{equation}
    X_{\text{LLM}} = [\mathcal{F}; E_{\mathcal{P}}] \in \mathbb{R}^{(N_q + L_{\mathcal{P}}) \times d}.
\end{equation}
Rather than predicting discrete labels directly, we formulate the task as prompt-based generation to better align with the LLM’s pretraining objective.
Verbalizers map each class to semantically meaningful tokens, improving generalization and interpretability, especially under label sparsity.
To adapt the LLM efficiently, we apply Low-Rank Adaptation (LoRA)~\cite{DBLP:conf/iclr/HuSWALWWC22}, injecting trainable updates $\Delta W = BA$ into the frozen weights $W_0$. The final prediction is made as:
\begin{equation}
    \begin{aligned}
        h &= \mathcal{M}_{\text{LLM}}(X_{\text{LLM}}; W_0) + \Delta W \cdot X_{\text{LLM}}, \\
    \hat{y} &= \arg\max_{v \in \mathcal{V}} \langle h_{-1}, E(v) \rangle,
    \end{aligned}
\end{equation}
where $h_{-1}$ denotes the last token's hidden state.
This prompt-based formulation leverages the LLM’s semantic priors to interpret behavioral nuances, while LoRA enables efficient task-specific tuning for insider threat detection with minimal parameter overhead.

Formally, the Insight-LLM procedure is outlined in Algorithm~\ref{algorithm}. 
Given user behavioral logs $\mathcal{L}$, a prompt $\mathcal{P}$, verbalizer tokens $\mathcal{V}$, view-specific encoders $\mathcal{E}$, and a LLM $\mathcal{M}_{LLM}$, the process proceeds in four stages. 
First, for each behavioral view $v \in \{\text{text}, \text{sent}, \text{seq}, \text{topo}\}$, a corresponding encoder $\mathcal{E}^{(v)}$ is used to extract features $\mathcal{X}^{(v)}$ (in line 7). 
Next, intra-view alignment projects and normalizes each $\mathcal{X}^{(v)}$ (in line 10), applies self-attention with learnable queries $Q^{(v)}$ (in line 11), and refines the result through residual and feed-forward layers to obtain the aligned representation $\tilde{Z}^{(v)}$ (in lines 12--13). 
These aligned features are then concatenated and passed to an inter-view fusion module via multi-head attention using cross-view queries $Q_{\text{inter}}$, resulting in the fused embedding $\mathcal{F}$ (in lines 16--17). 
Finally, the model tokenizes the prompt $\mathcal{P}$ (in line 19) and concatenates it with $\mathcal{F}$ as input to the LLM $\mathcal{M}_{LLM}$ (in line 20); the final hidden state $h_{\text{final}}$ (in line 21) is compared against class verbalizers $\mathcal{V}_c$ using cosine similarity to predict the threat label $\hat{y}$ (in line 22). 
This formulation enables the model to integrate structured behavioral signals with semantic priors from the LLM for enhanced threat detection.

\section{Experimental Setup}

\subsection{Research Questions}
\label{RQs}
We list several research questions (RQs) to guide the experiments and verify the effectiveness of our proposal.
\begin{enumerate}[nosep, align=left, leftmargin=*]
	\item[\bf{RQ1}:] Does the proposed Insight-LLM outperform state-of-the-art baselines on log-based insider threat detection?  
        \item[\bf{RQ2}:] Which components contribute most to improving model performance?  
        \item[\bf{RQ3}:] How does the multi-view fusion framework adjust view contributions across diverse threat scenarios?  
        \item[\bf{RQ4}:] How sensitive is the LLM-based fusion framework to parameter choices in adaptation modules, such as the LoRA rank and scaling factor?  
        \item[\bf{RQ5}:] How does the framework perform under different prompt and label word designs?  
\end{enumerate}

\subsection{Baseline Methods}
\label{baseilnes}


To rigorously evaluate the performance of Insight-LLM, we compare it with nine representative and competitive baselines spanning a wide spectrum of fusion strategies and single-view approaches. Fusion-based methods include 
shallow models (LaAeb~\cite{DBLP:journals/compsec/FeiZZGFLWC25}), 
text-oriented frameworks (ITDLM~\cite{DBLP:conf/coling/SongZG25}), 
image-driven designs~\cite{DBLP:journals/compsec/DuanYW24}, 
transformer-based architectures (LAN~\cite{DBLP:journals/tifs/CaiWXLZLY24}, ITDBERT~\cite{DBLP:conf/iscc/HuangZLLWY21}), and graph-based approaches (LMTracker~\cite{DBLP:conf/iscc/HuangZLLWY21}).
For single-view evaluation, we select representative models from each modality to ensure fair comparison, including Log2Vec~\cite{DBLP:conf/ccs/LiuWZJXM19} for topological features, GRU~\cite{DBLP:journals/corr/ChungGCB14} for sequential patterns, RoBERTa~\cite{DBLP:journals/corr/abs-1907-11692} for textual semantics, and TweetEval~\cite{DBLP:conf/emnlp/BarbieriCAN20} for sentiment analysis. 
This selection ensures that the baselines span both classical and state-of-the-art designs, providing a comprehensive and robust benchmark for assessing the effectiveness of Insight-LLM.

\subsection{Datasets}
\label{datasets}

\begin{table}[]
\caption{Statistics of the datasets. The imbalance ratio is calculated by $N_{maj}/N_{min}$, where $N_{maj}$ and $N_{min}$ are the sample sizes of the majority class and the minority class.}
    \begin{tabular}{lrr}
    \hline
    Dataset                       & \multicolumn{1}{l}{CERT r4.2} & \multicolumn{1}{l}{CERT r5.2} \\ \toprule
    \# Benign employees           & 930                           & 1,901                         \\
    \# Malicious employees        & 70                            & 99                            \\
    \# Normal activities          & 32,762,906                    & 79,846,358                    \\
    \# Abnormal activities        & 7,316                         & 10,306                        \\ \hline
    Imbalance Ratio of activities & 4,478                         & 7,748                         \\ \bottomrule
    \end{tabular}
    \
    \label{statistics}
\end{table}

Following prior work~\cite{DBLP:journals/tnsm/LeZH20,DBLP:journals/tifs/LiLJLYGY23,DBLP:conf/ccs/LiuWZJXM19,DBLP:conf/cikm/YuanZWT20,DBLP:journals/corr/abs-2408-08902}, we evaluate Insight-LLM on two publicly available datasets, CERT r4.2 and CERT r5.2~\cite{DBLP:conf/sp/GlasserL13}. 
Both datasets record user activities within a company from January 2010 to June 2011, but differ in scale. 
As summarized in Table~\ref{statistics}, CERT r4.2 contains 1,000 employees and 32,770,222 activities, including 7,316 abnormal activities manually injected for 70 employees by domain experts. CERT r5.2 expands to 2,000 employees and 79,856,664 activities, with 10,306 abnormal activities for 99 employees. In both datasets, normal activities dominate overwhelmingly.

Since user activity logs come from multiple sources (e.g., logon, website visits), we first aggregate them by user and timestamp. Following~\cite{DBLP:journals/tifs/CaiWXLZLY24}, each user’s activities are then split into sessions, defined as chronological sequences between logon and logoff.
Since the goal is to detect insider threats at the activity level, each activity is regarded as a sample for training the model.
Additionally, we use the user activity data in 2010 for training and validation, and use the data from January 2011 to June 2011 for evaluating all methods.

\begin{table*}[t]
\begin{spacing}{0.9}
\centering
\caption{Performance comparison of Insight-LLM with six baselines for insider threat detection. 
The best and second-best results are boldfaced and underlined, respectively. 
$\uparrow$: higher is better, $\downarrow$: lower is better.}
\label{overall_performance}
\renewcommand{\arraystretch}{1.15}
\begin{tabularx}{\textwidth}{ll *{8}{Y}}
\toprule
\multirow{2}{*}{\textbf{Category}} & \multirow{2}{*}{\textbf{Method}} 
& \multicolumn{4}{c}{\textbf{CERT r4.2}} 
& \multicolumn{4}{c}{\textbf{CERT r5.2}} \\ 
\cmidrule(lr){3-6} \cmidrule(lr){7-10}
& & \textbf{Prec}$\uparrow$ & \textbf{DR}$\uparrow$ & \textbf{FPR}$\downarrow$ & \textbf{F1}$\uparrow$ 
  & \textbf{Prec}$\uparrow$ & \textbf{DR}$\uparrow$ & \textbf{FPR}$\downarrow$ & \textbf{F1}$\uparrow$ \\ 
\midrule
\multirow{5}{*}{\textit{Single-view}} 
& GRU        & 0.8531 & 0.8032 & 0.1996 & 0.8392 & 0.8427 & 0.8131 & 0.1359 & 0.8221 \\
& TS2Vec     & 0.8710 & 0.7800 & 0.2013 & 0.8155 & 0.8436 & 0.7454 & 0.2419 & 0.7934 \\
& Node2Vec   & 0.9154 & 0.7987 & 0.7333 & 0.8530 & 0.8835 & 0.7549 & 0.7615 & 0.8249 \\
& RoBERTa    & 0.8988 & 0.8437 & 0.2569 & 0.8632 & 0.8746 & 0.8258 & 0.2696 & 0.8441 \\
& TweetEval  & 0.8786 & 0.8341 & 0.3349 & 0.8597 & 0.8452 & 0.7942 & 0.3663 & 0.8144 \\
\midrule
\multirow{6}{*}{\textit{Multi-view}} 
& Transformers & 0.7815 & 0.8005 & 0.2112 & 0.7946 & 0.7992 & 0.7958 & 0.2386 & 0.8259 \\
& LMTracker    & 0.7931 & 0.8824 & 0.1032 & 0.8961 & 0.7546 & 0.8418 & 0.1532 & 0.8637 \\
& ITDLM        & 0.8825 & 0.9171 & 0.1276 & 0.9141 & 0.8627 & 0.8957 & 0.1458 & 0.8842 \\
& ITDBERT      & 0.7354 & 0.6911 & 0.3153 & 0.7246 & 0.7159 & 0.6853 & 0.1996 & 0.7045 \\
& LAAEB        & 0.9153 & 0.8934 & 0.1072 & 0.9002 & 0.8972 & 0.8791 & 0.1244 & 0.8815 \\
& LAN          & \underline{0.9258} & \underline{0.9478} & \underline{0.1222} & \underline{0.9492} & \underline{0.9349} & \underline{0.9024} & \underline{0.0865} & \underline{0.9156} \\
\midrule \midrule
\multirow{2}{*}{\textit{Ours}} 
& Insight-LLM      & \textbf{0.9631} & \textbf{0.9683} & \textbf{0.0476} & \textbf{0.9712} 
               & \textbf{0.9512} & \textbf{0.9466} & \textbf{0.0496} & \textbf{0.9594} \\
& \textit{Improvements} & \textcolor{green!70!black}{4.0\%} & \textcolor{green!70!black}{2.2\%} & \textcolor{green!70!black}{61.0\%} & \textcolor{green!70!black}{2.3\%} & \textcolor{green!70!black}{1.7\%} & \textcolor{green!70!black}{4.9\%} & \textcolor{green!70!black}{42.7\%} & \textcolor{green!70!black}{4.8\%} \\
\bottomrule
\end{tabularx}
\label{overall_performance}
\end{spacing}
\end{table*}

\subsection{Experimental Configuration}
\label{configuration}

The experiments are conducted with an Intel Xeon(R) Gold 5218R CPU, 256 GB of RAM, and four Nvidia RTX A6000 (48 GB) GPUs.
All experiments use Meta-Llama-3-8B~\cite{openlm2023openllama,touvron2023llama} as the backbone (hidden size $d=4096$), with LoRA ($r=8$, $\alpha=16$, dropout $0.05$). 
Each ViewAdapter employs $M=4$ query tokens and intra-view multi-head attention with $8$ heads; inter-view fusion uses $N_q=8$ learnable fusion tokens and a $8$-head MultiheadAttention. The adapter hidden size is $768$. 
Training uses batch size $16$, $30$ epochs, AdamW (learning rate $5\times10^{-5}$, weight decay $0.01$) with StepLR (step size $2$, gamma $0.5$). 
Unless noted, we optimize cross-entropy with inverse-frequency class weights, and the prompt ``Classify user behavior types:'' with label-word verbalizers.

\section{Results and Analysis}

\subsection{Overall Performance}
\label{overall}

To answer \textbf{RQ1}, we evaluate the performance of our proposed Insight-LLM and six competitive baselines for the insider threat detection task.
We present the results in Table~\ref{overall_performance}. Higher Precision, Detection Rate (DR), and F1 indicate better performance, while lower FPR reduces false alarms.

Insight-LLM consistently outperforms the strongest baseline (LAN) across both datasets. 
On CERT r4.2, Precision improves from 0.9258 to 0.9631 (+4.0\%), DR from 0.9478 to 0.9683 (+2.2\%), F1 from 0.9492 to 0.9712 (+2.3\%), and FPR decreases sharply from 0.1222 to 0.0476 (-61.0\%). 
On CERT r5.2, Precision rises from 0.9349 to 0.9512 (+1.7\%), DR from 0.9024 to 0.9466 (+4.9\%), F1 from 0.9156 to 0.9594 (+4.8\%), and FPR drops from 0.0865 to 0.0496 (-42.7\%). 
These improvements indicate that Insight-LLM not only increases the ability to detect insider threats correctly but also substantially reduces false alarms, which is critical for operational deployment.
Analyzing the trends across datasets, Insight-LLM maintains stable performance: although CERT r5.2 is more challenging, FPR remains below 5\%, and DR stays above 94\%, showing robustness to dataset variations. 
Compared with single-view methods (e.g., RoBERTa or GRU), Insight-LLM achieves significantly higher DR and lower FPR, confirming that multi-view fusion effectively leverages heterogeneous signals to improve discrimination between normal and malicious behaviors.
Furthermore, the relative improvements in FPR are particularly notable: Insight-LLM reduces false positives by 42–61\% compared with LAN, while also improving Precision and DR. This demonstrates that the model not only captures more true positives but also avoids unnecessary alarms, striking a better balance between sensitivity and specificity. Overall, Insight-LLM provides substantial gains in both accuracy and reliability, highlighting the value of integrating multi-view representations for insider threat detection.

\subsection{Ablation Study in Insight-LLM}
\label{ablation}

\begin{table*}[t]
\centering
\caption{Ablation study results of Insight-LLM on CERT r4.2 and CERT r5.2. The biggest drop in each column is appended with $\downarrow$.}
\label{tab:ablation}
\begin{tabular}{clcccccccc}
\toprule
\multicolumn{2}{c}{\multirow{2}{*}{\textbf{Model Variants}}} 
& \multicolumn{4}{c}{\textbf{CERT r4.2}} 
& \multicolumn{4}{c}{\textbf{CERT r5.2}} \\ 
\cmidrule(lr){3-6} \cmidrule(lr){7-10}
\multicolumn{2}{c}{} & \textbf{Prec} & \textbf{DR} & \textbf{FPR} & \textbf{F1} 
                     & \textbf{Prec} & \textbf{DR} & \textbf{FPR} & \textbf{F1} \\ 
\midrule
\multirow{4}{*}{\shortstack{\textit{View}\\\textit{level}}} 
& w/o Text              & 0.6753$\downarrow$ & 0.6934$\downarrow$ & 0.3622$\downarrow$ & 0.6755$\downarrow$ & 0.6593$\downarrow$ & 0.6757$\downarrow$ & 0.4779$\downarrow$ & 0.6651$\downarrow$ \\
& w/o Sequence          & 0.7972 & 0.8131 & 0.2508 & 0.7837 & 0.7749 & 0.7924 & 0.2903 & 0.7606 \\
& w/o Topology          & 0.8992 & 0.9149 & 0.0962 & 0.8847 & 0.8647 & 0.8911 & 0.1125 & 0.8761 \\
& w/o Sentiment         & 0.9407 & 0.9458 & 0.0582 & 0.9344 & 0.9298 & 0.9328 & 0.0663 & 0.9217 \\
\midrule
\multirow{3}{*}{\shortstack{\textit{View}\\\textit{fusion}}} 
& w/o Intra-view Adapt. & 0.8813 & 0.8889 & 0.1205 & 0.8846 & 0.8397 & 0.8459 & 0.1569 & 0.8419 \\
& w/o Inter-view Fusion & 0.8556 & 0.8719 & 0.0692 & 0.8676 & 0.8429 & 0.8557 & 0.0819 & 0.8537 \\
& w/o FNN Projector     & 0.9036 & 0.9106 & 0.0982 & 0.8981 & 0.8845 & 0.9011 & 0.1215 & 0.8984 \\
\midrule
\multirow{2}{*}{\shortstack{\textit{LLM}\\\textit{Tuning}}} 
& w/o MLP Head          & 0.8515 & 0.8557 & 0.0756 & 0.8536 & 0.8316 & 0.8457 & 0.0913 & 0.8428 \\
& w/ Random Verbalizer  & 0.9317 & 0.9389 & 0.0657 & 0.9217 & 0.9198 & 0.9203 & 0.0752 & 0.9223 \\
\midrule \midrule
\multicolumn{2}{c}{Insight-LLM (original)} 
 & 0.9631 & 0.9683 & 0.0476 & 0.9712 & 0.9512 & 0.9466 & 0.0496 & 0.9594 \\
\bottomrule
\end{tabular}
\end{table*}

For RQ2, we perform an ablation study to evaluate the marginal effect of each view on overall performance and to assess the contribution of each component of Insight-LLM. Results are shown in Table~\ref{tab:ablation}.

Removing the text view leads to the largest degradation. On r4.2 precision drops from 0.9631 to 0.6753, detection rate from 0.9683 to 0.6934, F1 from 0.9712 to 0.6755, while FPR rises from 0.0476 to 0.3622. Similar trends appear on r5.2, with FPR rising to 0.4779.
The sequence view is the next most important. On r4.2 precision decreases by 0.1659, detection rate by 0.1552, F1 by 0.1875, and FPR rises by 0.2032. On r5.2, precision and detection rate drop by about 0.17 and 0.15, while FPR increases by 0.2407.
Fusion and tuning modules also matter. Removing intra-view adaptation changes r4.2 precision and detection rate by about 0.08 and increases FPR by 0.07; removing inter-view fusion changes precision by 0.11 and FPR by 0.02. FNN projector and MLP head yield similar moderate changes. Using a random verbalizer has smaller but consistent negative effects, e.g., precision decreases by 0.0314 and FPR rises by about 0.02–0.03.
Topology and sentiment views contribute smaller improvements: removing topology changes precision and detection rate by around 0.06 and FPR by 0.05, while removing sentiment only shifts metrics by less than 0.04.
Overall, text and sequence provide the most critical information, while fusion and tuning modules ensure stability and low false alarms. Topology and sentiment offer incremental but useful gains.

\subsection{Weight allocation}
\label{Weight_allocation}
\begin{figure}
	\includegraphics[width=0.49\textwidth]{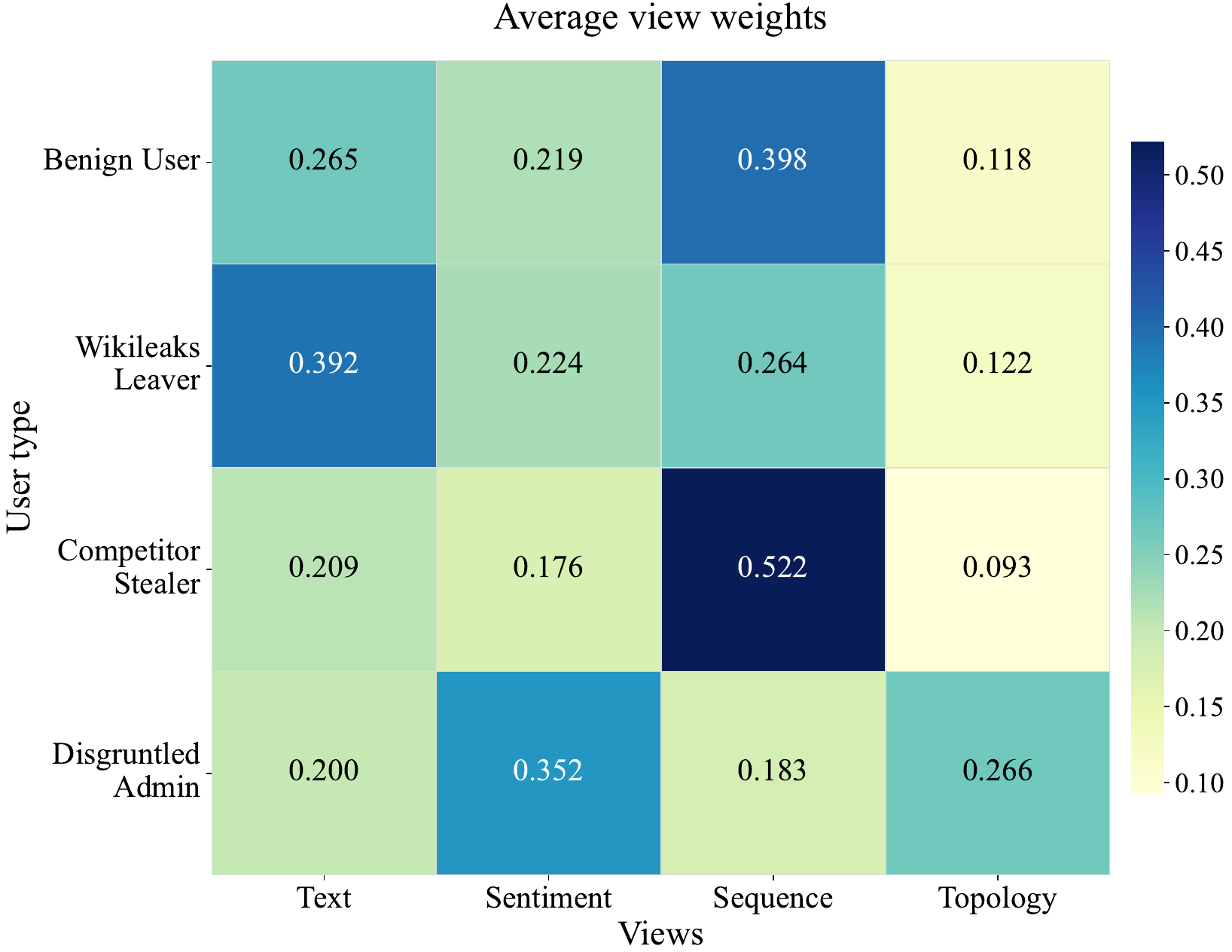}
	\caption{Heatmap of the average weights assigned to different behavioral views across user types.}
	\label{weight}
\end{figure}

To answer RQ3, we take CERT r4.2 as a case study and visualize how the model allocates modality weights across four representative categories of insider behaviour in Fig~\ref{weight}. 
For benign users, who operate entirely within normal routines without anomalous behaviour, the model relies mainly on sequence information (39.8\%), reflecting the importance of temporal regularities in routine activity. 
Wikileaks leavers, who log in after hours, use removable drives, upload data to wikileaks.org, and then resign, are distinguished by a strong emphasis on textual information (39.1\%), suggesting that content-based cues are crucial for leak-related behaviour. 
Competitor stealers, who solicit employment from rivals and substantially increase thumb-drive usage before departure, are characterized by sequence dominance (52.2\%), capturing the escalation of abnormal device activity. 
Disgruntled administrators, who install keyloggers on supervisors’ machines and misuse collected credentials to trigger panic before leaving, show higher weights on sentiment (35.2\%) and topology (26.5\%), indicating that affective signals and structural relationships are more informative in this case.

These results demonstrate that the model adapts modality allocations to the semantics of each class rather than distributing weights uniformly. Such class-specific patterns align with domain knowledge, though further validation via per-sample analysis and ablation remains necessary.

\vspace{-6pt}

\subsection{Parameter efficiency}
\begin{figure*}[t]
	\centering
	\begin{subfigure}[t]{0.32\textwidth}
		\includegraphics[width=\textwidth]{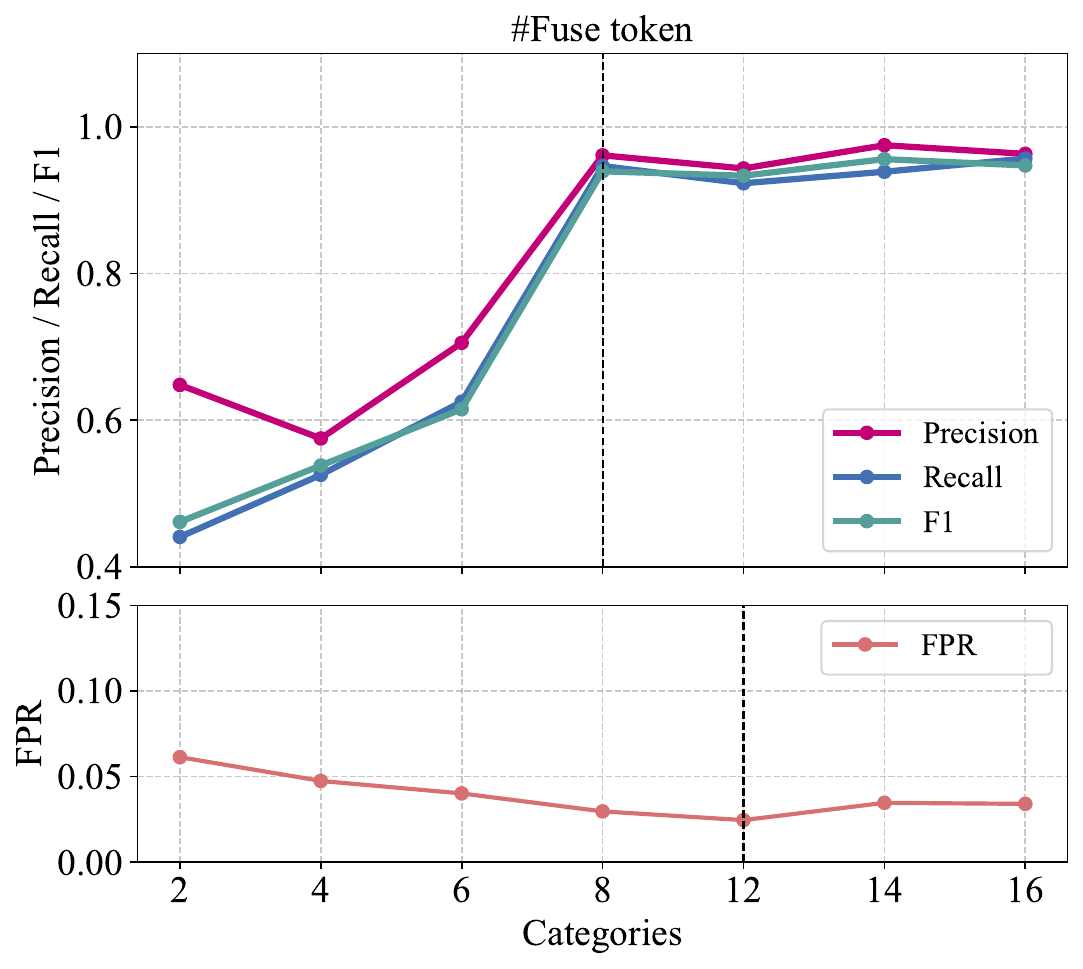}
		\vspace*{-1.5\baselineskip}
		\label{fuse_token}
	\end{subfigure}
	\begin{subfigure}[t]{0.32\textwidth}
		\includegraphics[width=\textwidth]{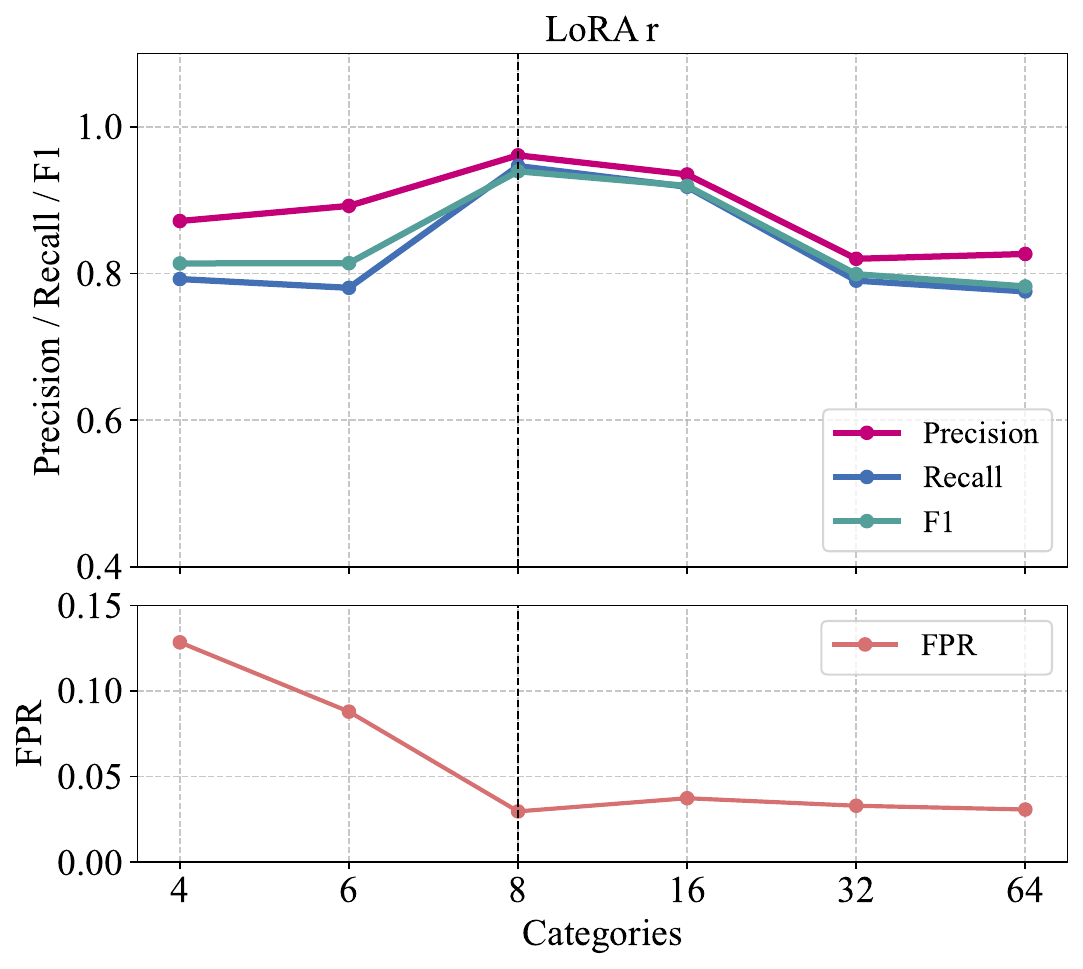}
		\vspace*{-1.5\baselineskip}
		\label{lora_r}
	\end{subfigure}  
	\begin{subfigure}[t]{0.32\textwidth}
		\includegraphics[width=\textwidth]{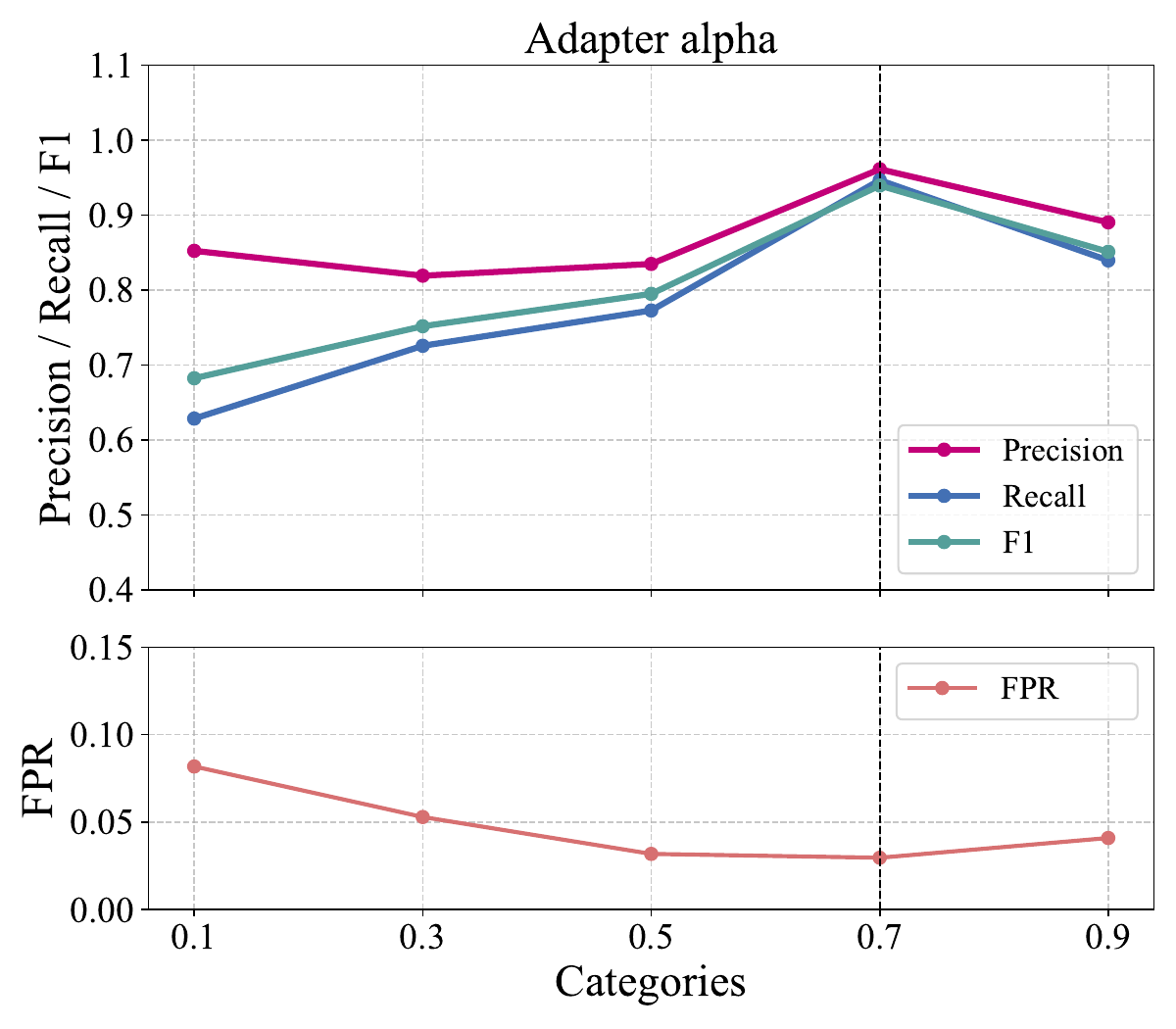}
		\vspace*{-1.5\baselineskip}
		\label{adapter_alpha}
	\end{subfigure}  
	\caption{Performance of Parse-LLM under different log characteristics.}
	\label{parameter_eff}
\end{figure*}

To answer RQ4, we conducted a controlled sensitivity study of Insight-LLM in which three single-factor sweeps were performed while holding other components fixed. 
The first sweep varied the number of fuse tokens, that is, the number of condensed tokens produced for each view, using the values $\{2,4,6,8,10,12,14\}$. 
The second sweep varied the LoRA rank r across the values $\{4,6,8,16,32,64\}$. 
The third sweep varied the adapter residual fusion weight $\alpha$ across the values $\{0.1,0.3,0.5,0.7,0.9\}$. 
For every configuration, we computed macro-averaged precision, recall, F1, and FPR. 
The following analysis refers to these validation metrics and identifies systematic trends that emerged from the three parameter families.

The results exhibit consistent trends across the three sets of parameters. For fuse tokens, performance improves rapidly when moving from two to six tokens, with F1 rising from 0.4613 to 0.6150 and FPR declining from 0.0613 to 0.0401. 
A marked gain appears at eight tokens (precision 0.9612, recall 0.9466, F1 0.9394), after which the model maintains stable and high performance for larger token counts. 
For LoRA rank, the results reflect a trade-off between representational capacity and generalization. Small ranks such as $r=4$ and $r=6$ achieve only moderate F1 (0.8138 and 0.8141) with relatively high FPR, while an intermediate rank of $r=8$ yields the best overall outcome (F1 0.9394, FPR 0.0296). 
Larger ranks retain competitive accuracy but eventually degrade when $r$ is pushed to 32 or 64, indicating overfitting. 
Finally, for adapter $\alpha$, the metrics improve steadily from $\alpha=0.1$ to $\alpha=0.7$, culminating in F1 0.9394 with FPR 0.0296, but performance diminishes at $\alpha=0.9$, where F1 drops to 0.8510. These patterns collectively highlight three mechanisms: insufficient cross-view representation at low fuse tokens, diminishing returns and overfitting at excessive LoRA ranks, and a balance between fusion reliance and noise amplification in the choice of $\alpha$.

\subsection{Impact of Base LLMs}
\label{base}

\begin{figure}[t]
	\centering
	\begin{subfigure}[t]{0.4\textwidth}
		\includegraphics[width=\textwidth]{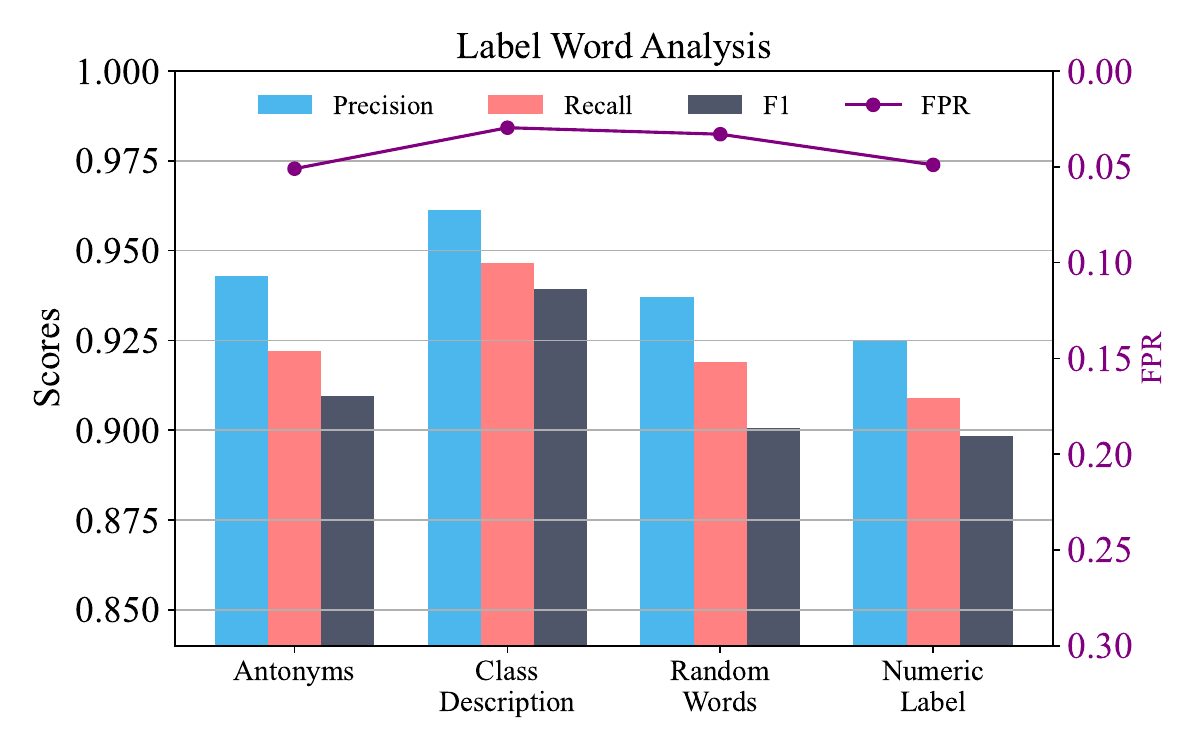}
		\caption{Performance on different label words.}
		\label{emb:distance1}
	\end{subfigure}
	\begin{subfigure}[t]{0.4\textwidth}
		\includegraphics[width=\textwidth]{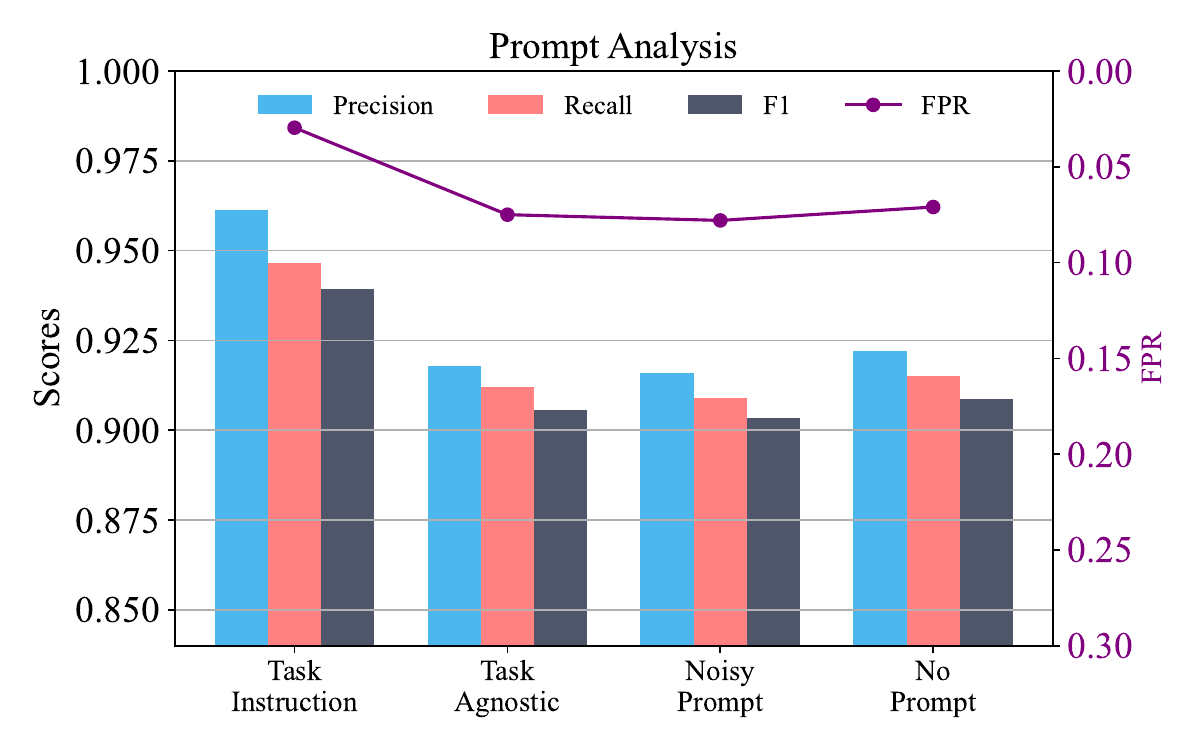}
		\caption{Performance on different prompt strategies.}
		\label{emb:distance2}
	\end{subfigure}  
    \caption{Performance comparison of different label word designs and prompt strategies.}
 \label{prompt_analyses}
\end{figure}

To address RQ5, we investigate how different prompt strategies and label word designs influence model performance. For label words, we consider four variants: 
\textit{Antonyms} (opposite terms of the ground-truth labels), 
\textit{Class Description} (semantic phrases describing the class meaning), 
\textit{Random Words} (arbitrarily selected words unrelated to the task), and 
\textit{Numeric Label} (numerical identifiers such as ``0'' or ``1''). 
For prompt strategies, we include 
\textit{Task Instruction} (explicit task descriptions), 
\textit{Task Agnostic} (generic prompts without task-specific information), 
\textit{Noisy Prompt} (irrelevant or misleading instructions), and 
\textit{No Prompt} (direct prediction without any prompting). 
The comparative results are summarized in Figure~\ref{prompt_analyses}.

For label word selection, the Class Description setting yields the best overall performance, achieving the highest precision, recall, and F1 score, while also maintaining the lowest false positive rate. 
In contrast, the Numeric Label strategy performs the worst across all metrics, with noticeably lower F1 score and a relatively high FPR.
This suggests that semantically meaningful label words provide stronger supervision signals, whereas arbitrary or numeric tokens hinder discriminative capability.
Regarding prompt strategies, incorporating Task Instruction leads to the most robust results, with F1 score of 0.9394 and the lowest FPR of 0.0296, clearly outperforming all other configurations. 
By comparison, removing task-specific guidance (No Prompt) or introducing irrelevant information (Noisy Prompt) causes performance degradation, reflected in lower precision and higher false positive rates (0.071-0.078). 
The Task Agnostic prompts show moderate effectiveness but still lag behind explicit instructions. These findings demonstrate that both prompt informativeness and semantically aligned label words are critical factors in optimizing detection accuracy while minimizing false alarms.

\section{Conclusion and future work}
In this study, we proposed Insight-LLM, a multi-view fusion framework for log-based insider threat detection. 
Insight-LLM employs dedicated ViewAdapters to map textual, sentiment, sequential, and topological features into a shared latent manifold, where intra-view cross-attention distills semantically meaningful tokens from each modality. 
An inter-view cross-attention mechanism with learnable fusion queries then integrates complementary signals across modalities, dynamically emphasizing features most indicative of each behavior type. 
The fused representations are combined with a natural language prompt and fed into a LoRA-adapted LLM, enabling behavior classification via verbalizer tokens while leveraging the LLM’s pretrained semantic knowledge. 
Experimental results on CERT datasets demonstrate that Insight-LLM effectively captures subtle and dispersed threat signals, outperforms single-view and traditional fusion baselines, and maintains robustness under severe data imbalance. 
The combination of structured multi-view alignment, cross-modal fusion, and prompt-based LLM prediction allows Insight-LLM to achieve interpretable and accurate insider threat detection.

\bibliographystyle{IEEEtran}
\bibliography{MV-ITD}

\end{document}